\documentclass[10pt,journal,compsoc]{IEEEtran}
\usepackage[noadjust,compress]{cite}
\usepackage[utf8]{inputenc}
\usepackage{balance}
\usepackage{graphicx}
\usepackage[utf8]{inputenc}
\usepackage[T1]{fontenc}
\usepackage[usenames, dvipsnames]{xcolor}
\usepackage{colortbl}
\definecolor{Gray}{gray}{.9}
\usepackage{pifont}
\usepackage{makecell}

\usepackage{tikz}
\newcommand\copyrighttext{%
  \tiny \textcopyright 2020 IEEE. Personal use of this material is permitted. Permission from IEEE must be obtained for all other uses, in any current or future media, including reprinting/republishing this material for advertising or promotional purposes, creating new collective works, for resale or redistribution to servers or lists, or reuse of any copyrighted component of this work in other works. Cite this article as follows: R. Ošlejšek, V. Rusnak, K. Burská, V. Švábenský, J. Vykopal and J. Čegan, \textit{Conceptual Model of Visual Analytics for Hands-on Cybersecurity Training}, in IEEE Transactions on Visualization and Computer Graphics. DOI: \url{https://doi.org/10.1109/TVCG.2020.2977336}.}
\newcommand\copyrightnotice{%
\begin{tikzpicture}[remember picture,overlay]
\node[anchor=south,yshift=12pt] at (current page.south) {\fbox{\parbox{\dimexpr\textwidth-\fboxsep-\fboxrule\relax}{\copyrighttext}}};
\end{tikzpicture}%
}

\begin{document}

\bstctlcite{IEEEexample:BSTcontrol}

\title{Conceptual Model of Visual Analytics for~Hands-on Cybersecurity Training}

\author{Radek~Ošlejšek, Vít~Rusňák, Karolína~Burská, Valdemar~Švábenský, Jan~Vykopal, and~Jakub~Čegan
\IEEEcompsocitemizethanks{\IEEEcompsocthanksitem R. Ošlejšek and K. Burská are with the Faculty of Informatics, Masaryk University, Brno, Czech Republic.\protect\\
E-mail: \{oslejsek, xburska\}@fi.muni.cz
\IEEEcompsocthanksitem V. Rusňák, and J. Čegan are with the Institute of Computer Science, Masaryk University, Brno, Czech Republic.\protect\\
E-mail: \{rusnak, cegan\}@ics.muni.cz
\IEEEcompsocthanksitem V. Švábenský and J. Vykopal are with the Institute of Computer Science and Faculty of Informatics, Masaryk University, Brno, Czech Republic.\protect\\
E-mail: \{svabensky, vykopal\}@ics.muni.cz}
\thanks{Manuscript received August 6, 2019.}}

%

\IEEEtitleabstractindextext{%
\begin{abstract}
Hands-on training is an effective way to practice theoretical cybersecurity concepts and increase participants’ skills. In this paper, we discuss the application of visual analytics principles to the design, execution, and evaluation of training sessions. We propose a conceptual model employing visual analytics that supports the sensemaking activities of users involved in various phases of the training life cycle. The model emerged from our long-term experience in designing and organizing diverse hands-on cybersecurity training sessions. It provides a classification of visualizations and can be used as a framework for developing novel visualization tools supporting phases of the training life-cycle. 
We demonstrate the model application on examples covering two types of cybersecurity training programs. 
\end{abstract}

\begin{IEEEkeywords}
Visual analytics, cybersecurity, hands-on training, classification, education.
\end{IEEEkeywords}}

\maketitle
\copyrightnotice

\IEEEdisplaynontitleabstractindextext

\IEEEraisesectionheading{\section{Introduction}\label{sec:intro}}

Our society is being exposed to an increasing number of cyber threats and attacks. The lack of a strong cybersecurity workforce presents a critical danger for companies and nations~\cite{cyberjobs2015}. 
Hands-on training of new professionals is an effective way to remedy this situation. In our work, we use visual-based sense-making and reasoning to support participants in better and faster comprehension of attacks, threats, and defense strategies. 

The ability to use visual-based analytical reasoning is essential in many fields, including biology~\cite{Krone2016}, medicine~\cite{lawonn2018}, urbanization~\cite{huang2016}, and education~\cite{govaerts2012}. The goal of this paper is to create a conceptual framework providing broader insight into the application of visual analytics (VA) principles~\cite{wong04} in hands-on cybersecurity training. Conceptual models like the one proposed in this paper help researchers design effective visual techniques in a given domain. To the best of our knowledge, the current literature for cybersecurity training lacks such a conceptual model.

There are several reasons for the absence of a conceptual model. Existing hands-on cybersecurity training is largely heterogeneous. Training sessions differ in content, organization, target audience, and technical means. Moreover, the cybersecurity domain represents a sensitive area similar to military or intelligence services, in which many sources are secret or restricted. Therefore, it is challenging to become familiar with this domain and clarify the terms and processes. Fortunately, we have the benefit of seven years of experience with the design and organization of training sessions. The results of this paper arise from close cooperation with domain experts who directly participate in the development and operation of the \emph{KYPO Cyber Range}~\cite{vykopal2017kypo} -- a sophisticated platform for cybersecurity training. Their knowledge and the survey of other existing approaches are essential for this work.

The two most widely recognized hands-on cybersecurity training activities are \emph{Capture the Flag} (CTF) and the \emph{Cyber Defense Exercise} (CDX). The main difference lies in their educational goals. While CTFs focus mainly on improving hard skills in the cybersecurity domain, CDXs target both hard and soft skills. CTF features a game-like approach~\cite{vigna2014, davis2014, werther2011, doupe2011}. Participants gain points for solving technical tasks that exercise their cybersecurity skills. Completing each task yields a text string called \emph{flag}. In contrast, CDXs have been traditionally organized by military and governmental agencies~\cite{petullo2016} that emphasize realistic training scenarios that authentically mimic the operational environment of a real organization~\cite{eagle2013}. We deeply analyzed these types of training programs to distill a unified visual analytics model that fits the heterogeneous cyber-training events and is simultaneously instructive for the design of specialized visual analytics tools. 

The major contributions of this paper are: 
(a) a definition of a unified training life cycle with user roles having clear responsibilities and requirements;
(b) a proposal for a conceptual model of visual analytics for hands-on cybersecurity training that can be used as a framework for further research and for developing visualizations supporting particular life-cycle tasks; and
(c) demonstrations of the applicability of the model using real examples and lessons learned from our long-term experience in designing and organizing hands-on cybersecurity training.

The paper is organized as follows: Section~\ref{sec:relate-work} introduces the related work. In Section~\ref{sec:life-cycle}, we discuss the generic life cycle of hands-on cybersecurity training sessions with user roles that delimit requirements put on analytical tasks and visualizations.
Sections~\ref{sec:data} and~\ref{sec:viz} provide classification schemes for data and analytical visualizations. A demonstration of the conceptual model is presented in Section~\ref{sec:use-cases}. Section~\ref{sec:discussion} summarizes the observations attained during our research. Section~\ref{sec:conclusion} outlines the direction for future research topics.

\section{Related Work} 
\label{sec:relate-work}

Our work is unique in its close interconnection of three areas: visual analytics, cybersecurity, and education. Publications dealing directly with the intersection of these fields are rare. Therefore, we have explored related work from several relevant points of view.

\subsection{Visual Analytics in Cybersecurity}

Many works have addressed the challenges related to the design or evaluation of cybersecurity tools and techniques~\cite{staheli14,best14,attipoe16,damico16,adams2018}. A visual analytics approach to automated planning attacks has been discussed \cite{yuen2015}. All the surveys have confirmed the importance of supporting analytic tasks by visual interfaces. However, they are aimed at the security-related focus only and do not tackle the educational aspect of the training of new experts. We took the challenges into account in our work, and we incorporated specific aspects of hands-on cybersecurity exercises. 

\subsection{Visual Analytics in Education and Training}

Another perspective that considers visualizations in relation to cybersecurity emphasizes the educational aspect. There are distinct approaches to enhancing cybersecurity abilities that focus on training or teaching computer security~\cite{schweitzer09,yuan10,fouh12}. However, these works again provide outputs of a narrow scope and often omit any profound conceptualization of their findings.

To help us comprehend the topic more thoroughly, we do not focus exclusively on the cybersecurity field; we also consider studies that relate to education and training from a broader view. 
A recent survey~\cite{firat18} introduces a literature classification in the field of interactive visualization for education with a focus on evaluation, and it lists common categories of educational visualizations from distinct fields. In this respect, our work is unique as it considers more than the educational theory. It also includes the application of hands-on training with practical and technical aspects that are an essential part of the learning process.  

The issue of education has been approached from the opposite direction \cite{macfadyen10}. In this work, the authors focus on predictive models for teachers of higher education institutions. They confirm the need for insight for both the teachers and the students that exceed simple summative feedback. 

\subsection{Generic Models of Visual Analytics}

Many generic design frameworks, models, and methods exist in the literature. These provide a structure and explanation of activities that designers perform when proposing suitable visualization tools~\cite{endsley2016,mckenna2015,sedlmair2012,koh2011}. However, the aim of this paper is not to discuss processes leading to the development of specific visualizations for cybersecurity training. Instead, we provide a conceptualization of the domain so that our model can serve as a framework for discussion and the efficient application of existing design methods for specific training tasks.

\begin{figure}[ht]
  \centering
  \includegraphics[width=\columnwidth]{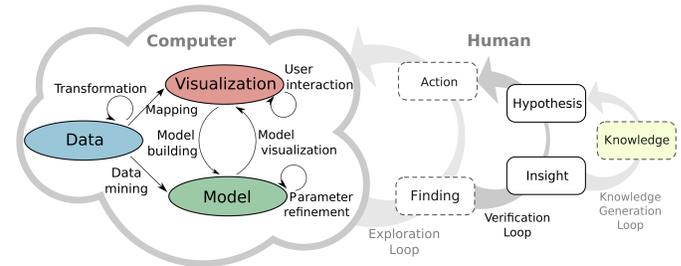}
  \caption{Altered version of models by Keim~\cite{keim} and Sacha~\cite{sacha} for insight retrieval based on visual analytics approaches.} 
  \label{fig:kg-model}
\end{figure}

Our solution builds upon Keim's~\cite{keim} and Sacha's~\cite{sacha} conceptual models for the visual analytics process. The VA process is characterized by the interaction between data, visualizations, models of the data, and users discovering knowledge, as shown in Fig.~\ref{fig:kg-model}. Keim emphasizes the computer-driven components of the VA process; Sacha extends the model with human reasoning. \emph{Data} carries facts in structured, semi-structured, or unstructured form. The \emph{model} captures the results of automated analysis methods. The interactive \emph{visualizations} are the primary user interface presenting \emph{data} and \emph{models} in a comprehensible manner. The human-centered part consists of three loops. The \emph{exploration loop} captures low-level visual interactions using actions and findings that are specific for individual visualizations and interests. The analysts then refine their hypotheses in the \emph{verification loop}. The \emph{knowledge generation loop} describes the transition from observations into generalized knowledge.

These two models form the foundations of our work. We utilize \emph{data} and \emph{visualization} components of Keim's model and narrow our focus on the \emph{verification loop} that plays a crucial role in building knowledge in any domain. The \emph{model} component of the VA process represents the cross-cutting concern, which is out of the scope of this paper. Therefore, we do not provide a separate classification for it. Instead, we mention suitable \emph{models} in our discussion of the classification of \emph{visualizations} and \emph{hypotheses}. The \emph{exploration loop} and \emph{knowledge generation loop} are omitted since they provide either too detailed or too generic concepts.

\section{Cybersecurity Training Life Cycle}
\label{sec:life-cycle}

The human loops of Sacha's VA model (see Fig.~\ref{fig:kg-model}) reflect the needs of users who interact with the computer system. Based on the literature review, our experience, and the application of analytical methods, we distilled the following general life cycle that clarifies \emph{who} is involved in the human loops, \emph{what} they expect (at a high level of abstraction), and \emph{when} they conduct their VA tasks. These pieces of information are later used for the detailed conceptualization of the ``computer part'' of the VA model by answering \emph{what} (data and hypotheses) and \emph{how} (visualizations) can be analyzed in the cyber training.

\subsection{Phases}

\begin{figure}[!ht]
  \centering
  \includegraphics[width=.7\columnwidth]{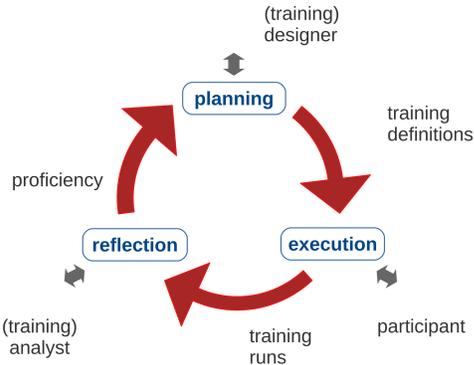}
  \caption{Cybersecurity training life-cycle phases with corresponding user roles, and main outcomes of each phase.}
  \label{fig:life-cycle}
\end{figure}

Based on the literature review and our experience, we distilled three generic phases (see Fig.~\ref{fig:life-cycle}) of the cybersecurity training life cycle. We performed a theory-driven qualitative coding method~\cite{saldana2015} on four key papers~\cite{vykopal2017, CJCSM2012, MITRE2014, HITRUST2015} that deal with organizational aspects of cybersecurity training. Using an open coding method helped us to structure the analysis and consolidate observations. Phases and outcomes discussed in the analyzed papers can slightly differ from our model. Nevertheless, the subtleties are rather negligible since the terminology in this domain is yet not established.

{\bf Planning} is the first phase of any new training. The goal is to formulate technical and educational requirements, set measurable objectives, and allocate necessary resources. The {\em training definition} -- the main output -- is a set of (more or less) formally defined configurations of the computer network and its nodes, specification of attacks, training tasks and objectives, scoring rules, expected skills of participants, and related configuration data of the training.

The {\bf execution} phase represents a training session in which participants are physically involved. User activities and the state of the training infrastructure are monitored, and the data is stored for further analysis. We refer to the data from this phase as {\em training runs}. 

During the {\bf reflection} phase, \emph{training definitions} and \emph{training runs} are analyzed and evaluated. Reflection can be conducted at any time. \emph{Analysts} usually explore the data after each training run to learn from it or provide feedback to involved people. However, they can also analyze the data before or during the planning phase of a new training session to gradually improve its quality. The reflection phase, therefore, helps to increase the \emph{proficiency} in designing and organizing training events.

\subsection{User Roles}
\label{sec:roles}

The requirements put on visual analytic interfaces are affected by user roles. The basic roles emerged from the life cycle. They reflect individual phases captured in Fig.~\ref{fig:life-cycle}. For clarity, our roles are \textsc{capitalized} in the paper.

\textsc{\textbf{Training designers}} (\textsc{\textbf{designers}} for short) are responsible for the design of training definitions during the \emph{planning} phase. Multiple designers with different skills are usually involved in the preparation of new training content. Cybersecurity experts contribute primarily to the technical aspects; education experts are responsible for defining the learning objectives and assessment criteria.

\textsc{\textbf{Participants}} represent everyone involved in the training event. Their analytical activities are associated with situational awareness and gaining insight into the training during the \emph{execution} phase.

The \textsc{\textbf{training analyst}} (\textsc{\textbf{analyst}} for short) role covers all the people who conduct the post-training analysis of collected data. In our VA model, this role is used to capture the requirements of generic analytical interactions. Various people interested in the relevant data can take on this role, e.g., cybersecurity experts looking for talented participants.

\begin{figure}[!ht]
  \centering
  \includegraphics[width=.85\columnwidth]{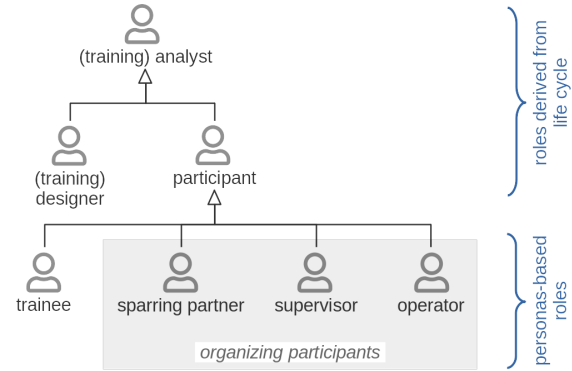}
  \caption{Hierarchy of user roles participating in cybersecurity training.}
  \label{fig:roles}
\end{figure}

These three roles are not independent. Arrows in Fig.~\ref{fig:roles} represent the inheritance of user roles as defined by requirements analysis methodologies in software engineering~\cite{schneider2001}. It means that \textsc{designers} and \textsc{participants} can conduct post-training analysis like other \textsc{training analysts}, e.g., to get feedback on completed training sessions. On the other hand, they can have a specific responsibility during the \emph{planning} or \emph{execution} phases, respectively.

The high-level roles that emerged from the life cycle proved to be too general to capture the fine-grained requirements of heterogeneous groups of people participating in real training events. Therefore, we employed the personas design method~\cite{hanington2012} to reveal archetypal users and further decompose user roles. We analyzed the same sources that we used during the conceptualization of the life cycle~\cite{vykopal2017, CJCSM2012, MITRE2014, HITRUST2015}. The observed personas are summarized in Table~\ref{tab:roles}.

CTF training includes only two types of personas, which correspond to a teacher-student relation. The \textit{student} (or \textit{learner}) follows instructions defined by the \emph{training definition} and performs the required tasks. The \textit{instructor} facilitates the training session from the educational point of view. Moreover, the instructor is also responsible for the technical aspects of training and addresses any possible technical difficulties with the underlying infrastructure.

In CDXs, we identified seven personas. \textit{Blue team} members are similar to \textit{learners} of CTFs. They have to defend the entrusted network from the attacks of the \textit{red team}. \textit{White team} members are responsible for the organization and compliance with the ``game rules'' of a CDX. \textit{Fictitious users} represent common users of the defended network. \textit{Law enforcement officers} check whether the actions of the \textit{blue team} are legal. \textit{Journalists} request reports from the \textit{blue teams}. Finally, the \textit{green team} is responsible for maintaining the infrastructure of the exercise.

By deeply analyzing the responsibilities and analytical goals of identified personas, we generalized them to four user roles. The mapping is captured in Table~\ref{tab:roles}.

\begin{table}[!ht]
\caption{Mapping of CTF/CDX personas to fine-grained user roles.}
\label{tab:roles}
\centering
\begin{tabular}{l|ll}
{\bf user roles} & {\bf CTF personas} & {\bf CDX personas} \\
\rowcolor{Gray}
\em trainee & student (learner) & blue team \\
\em sparring partner & -- & red team \\
   && white team \\
   && fictitious user  \\
   && law enforcement officer \\
   && journalist \\
\rowcolor{Gray}
\em supervisor & instructor & green team \\
 \rowcolor{Gray}
 && white team \\
\em operator & instructor & green team \\
\end{tabular}
\end{table}

\textsc{\textbf{Trainees}} solve tasks described in the \emph{training definition}. Their activities are monitored and assessed. They can work either individually or in teams. For the sake of simplicity, we use the term ``trainee'' for both cases.

\textsc{\textbf{Sparring partners}} represent individuals or teams involved in training sessions who actively compete with \textsc{trainees} but who are not directly assessed. Sparring partners also follow the instructions from the \emph{training definition}. However, their requirements for data analysis, feedback, and other educational aspects differ from the requirements for \textsc{trainees}. 

\textsc{\textbf{Supervisors}}, unlike \textsc{sparring partners}, do not follow the exact rules of the \emph{training definition}. They are responsible for overseeing the training session, enforcing rules, and other activities that are not exactly defined.

\textsc{\textbf{Operators}} are responsible for the underlying (technical) infrastructure of the hands-on training. This role requires technical skills and a good knowledge of the underlying technologies. The work of operators can significantly affect the course of the exercise since any technical difficulties can devalue educational results regardless of how well the training session has been prepared.

All the roles distilled from personas represent participants directly involved in a specific training session. Therefore, they are defined as descendants of the \textsc{participant} role in the schema in Fig.~\ref{fig:roles}. While \textsc{trainees} are the primary subject of training sessions, \textsc{sparring partners}, together with \textsc{supervisors} and \textsc{operators}, represent backstage \emph{organizing participants}.

\section{Data} 
\label{sec:data}

Visualizations designed for operational cybersecurity deal with large data sets~\cite{best14}. In contrast, training events are limited in time, resources, and the number of participants. As a result, the amount of data produced during the training sessions is also usually limited. However, the data is highly heterogeneous. Therefore, our classification has been developed iteratively together with the analysis of other parts of the VA model. The proposed scheme comes from the unified life cycle. Data categories reflect user roles and training phases during which the data is created. It enables us to clarify what data is available in each phase and define limitations to be considered in analytical visualizations. 

{\bf Technical scenarios (D\textsubscript{1})} capture the technical aspects and predefined processes of a \emph{training definition}. The technical aspects include, for example, the definition of the network topology, software running on individual network nodes (operating system, applications, services), and vulnerabilities injected in the network nodes. User procedures are defined as attack plans (attack vectors and their timing), \textsc{trainees'} tasks, hints, and other formalized steps.

{\bf Assessment criteria (D\textsubscript{2})} determine how to assess \textsc{trainees} and how to measure whether learning objectives were achieved. Assessment criteria define metrics, indicators, and aspects of the training related to the evaluation of \textsc{trainees}. Apart from that, the criteria can also include the definition of questionnaires for prerequisite testing of \textsc{trainees}, assessment questions during the exercise, and post-training feedback surveys. 

{\bf User actions (D\textsubscript{3})} are \textsc{participants'} actions monitored and collected during the \emph{execution} phase. Examples include commands entered by \textsc{trainees}, displayed hints, performed attacks or defenses and their results, intervention of \textsc{supervisors}, and other user-oriented events. 

{\bf Infrastructure data (D\textsubscript{4})} represent the state of computer networks and the underlying technical infrastructure. The data encodes node availability, available services, packet flows, and the health of the infrastructure. The obtained information can be used for direct infrastructure surveillance, and the assessment of \textsc{trainees} (e.g., \textsc{trainees} can be penalized for the unavailability of required services). 

{\bf Assessment data (D\textsubscript{5})} are related to the \emph{assessment criteria} and determine the success rate of \textsc{trainees} and their results in achieving learning objectives. The data encodes how successfully a particular user has solved a particular task (in percentages or as obtained penalties), time spent on tasks, answers to questionnaires, and other qualitative and quantitative indicators of the learning process. A great deal of quantitative data can be computed automatically by applying assessment criteria ($D_2$) to monitored user actions and infrastructure data ($D_3$ and $D_4$).

\begin{table}[!ht]
\centering
\caption{Data types mapping on life cycle phases, abstract data levels, and terminology from the paper.}
\label{tab:data}
\begin{tabular}{lll}
 & {\bf D\textsubscript{1} \& D\textsubscript{2}} & {\bf D\textsubscript{3} \&  D\textsubscript{4} \& D\textsubscript{5}} \\
\rowcolor{Gray}
{\em phase of creation} & planning & execution \\
{\em level of abstraction} & configuration data & operational data \\
\rowcolor{Gray}
{\em terminology} & training definition & training run
\end{tabular}
\end{table}

Mapping data categories to the planning and execution phases follows data abstraction as defined by Fowler for software systems~\cite{fowler1997}: $D_1$ and $D_2$ represent data from the \emph{configuration level}. They are defined during the \emph{planning} phase by \textsc{designers} as a part of \emph{training definitions}. $D_3$--$D_5$ represent data from the \emph{operational level}. They are acquired during the \emph{execution} phase and we refer to them as \emph{training runs}, as summarized in Table~\ref{tab:data}.

\section{Visualizations and Hypotheses} 
\label{sec:viz}

\begin{figure*}[!ht]
  \centering
  \includegraphics[width=\textwidth]{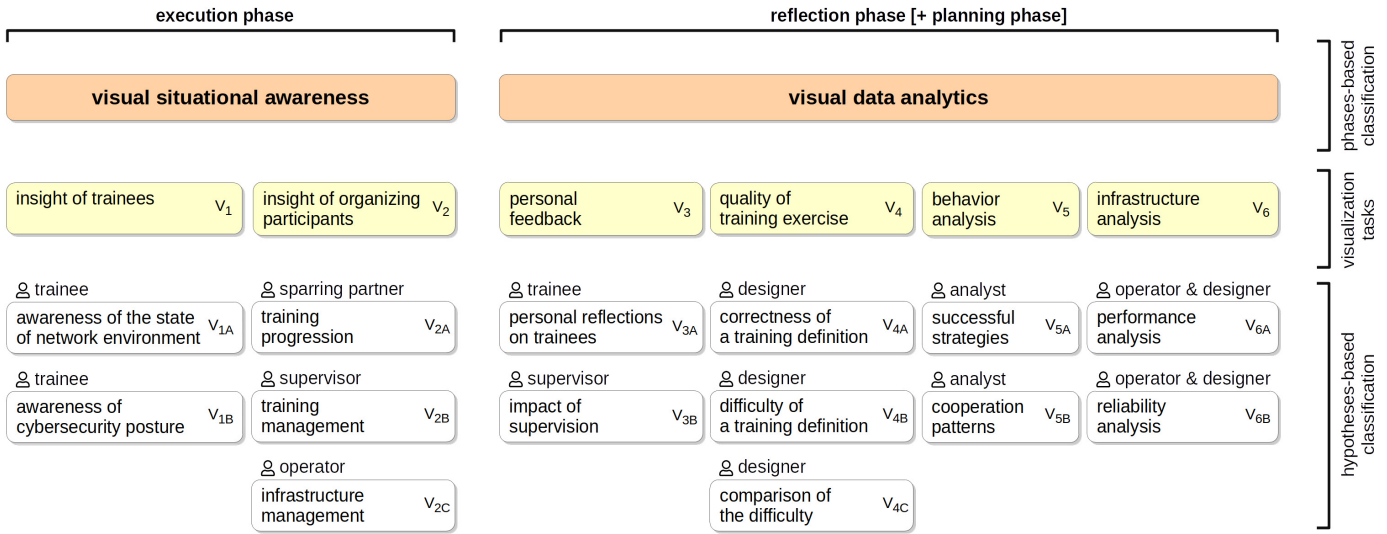}
  \caption{Classification of visualizations and hypotheses in the context of hands-on cybersecurity training.}
  \label{fig:classification}
\end{figure*}

According to the VA model of Sacha \& Keim (see Fig.~\ref{fig:kg-model}), requirements applied to visualizations are driven by hypotheses that people consider during their analytical activities. Therefore, we discuss and classify both visualizations and hypotheses together.

The classification shown in Fig.~\ref{fig:classification} was established iteratively by balancing two complementary directions. We broke down the top-level phases and roles of the training life cycle and, concurrently, we searched for low-level hypotheses that we organized into clusters. Balancing these two approaches, we concluded with a three-level classification scheme that, to the best of our knowledge, sufficiently covers the problem domain and emphasizes the design requirements of visual analytic tools. The low-level hypotheses were obtained from discussions with six domain experts (three of them are co-authors of this paper), each with more than six years of experience with organizing CTFs and CDXs. The final classification hierarchy was reached by consensus of the authors whose expertise includes cyber training design and organization as well as the design of analytical visualizations for KYPO Cyber Range~\cite{vykopal2017kypo}. The rest of this section is structured according to the proposed scheme as follows.

The top-level categories of \emph{Visual Situational Awareness} and \emph{Visual Data Analytics} in Fig.~\ref{fig:classification} represent distinct concepts using different data in different phases of the life cycle. They are discussed in two separate subsections. During conceptualization, we observed that the analytical tasks of \textsc{training designers} represent a subset of activities associated with the \emph{reflection} phase of \textsc{training analysts}. Hypotheses and visualizations of the \emph{planning} phase are, therefore, covered by the \emph{Visual Data Analytics} category.

Classification at the second level defines key visualization tasks V\textsubscript{1}--V\textsubscript{6} that are detailed later in this section. They differ in the roles involved in the visual analysis, analytical goals, and other aspects. Discussion is primarily focused on visual requirements and justification for the third-level classification of hypotheses V\textsubscript{1A}--V\textsubscript{6B}.

Providing an exhaustive list of hypotheses for each task V\textsubscript{1A}--V\textsubscript{6B} is impossible; they emerge continuously as users conduct analyses and gain insights into the solved problem. Instead, we discuss an abstraction used for the classification and propose several hypotheses as examples.

\subsection{Visual Situational Awareness}

Existing theoretical concepts of situational awareness distinguish between \emph{perception}, \emph{comprehension}, and \emph{projection} corresponding to the three levels of the well-known Endsley model~\cite{endsley1995}. However, the significance and meaning of the levels can differ in the context of cybersecurity training depending on users' roles and their goals. This is because providing comprehensive insight into cybersecurity events during the \emph{execution} phase can be undesirable in certain circumstances. This aspect is reflected in our classification, as discussed in what follows. Table~\ref{tab:hypotheses1} summarizes visualizations and hypotheses for situational awareness.

\begin{table}[!ht]
\centering
\caption{Visual Situational Awareness: Visualization tasks V\textsubscript{1} and V\textsubscript{2} are further divided into two (V\textsubscript{1A}--V\textsubscript{1B}), and three (V\textsubscript{2A}--V\textsubscript{2C}) categories. Each category is accompanied by sample hypotheses formulated as prerequisites for verification (``I suppose that \dots'').}
\label{tab:hypotheses1}
\begin{tabular}{|l|}
\hline
\rowcolor{Gray}
\multicolumn{1}{|c|}{\bf V\textsubscript{1} -- Insight of Trainees} \\
{\bf Awareness of the state of network environment (V\textsubscript{1A}):} \\
As a \emph{trainee}, I suppose that \dots \\
\hspace{1em} \dots the web running at host $X$ is accessible for users. \\
\hspace{1em} \dots the host $X$ is accessible for me via SSH. \\
\hspace{1em} \dots the external network (including internet) remains accessible. \\
{\bf Awareness of cybersecurity posture (V\textsubscript{1B}):} \\
As a \emph{trainee}, I suppose that \dots \\
\hspace{1em} \dots server $X$ I am defending is now under attack. \\
\hspace{1em} \dots my previous attack actions were successful. \\
\hspace{1em} \dots I have successfully protected server $X$ against \\ 
    \hspace{2.4em} the DDoS attack. \\
\hline
\rowcolor{Gray}
\multicolumn{1}{|c|}{\bf V\textsubscript{2} -- Insight of Organizing Participants} \\
{\bf Training progression (V\textsubscript{2A}):} \\
As a \emph{sparring partner}, I suppose that \dots \\
\hspace{1em} \dots the trainee $X$ completed task $Y$, a prerequisite for task $Z$. \\
\hspace{1em} \dots the DDoS attack against host $X$ defended by trainee $Y$ \\
    \hspace{2.4em} was successful. \\
\hspace{1em} \dots trainee $X$ fixed the vulnerability allowing a DDoS attack \\
    \hspace{2.4em} at host $Y$. \\
{\bf Training management (V\textsubscript{2B}):} \\
As a \emph{supervisor}, I suppose that \dots \\
\hspace{1em} \dots all trainees completed task $Y$, a prerequisite for task $Z$. \\
\hspace{1em} \dots trainee $X$ solved the task successfully. \\
\hspace{1em} \dots trainee $X$ is in trouble (working on task longer than $Y$ min). \\
{\bf Infrastructure management (V\textsubscript{2C}):} \\
As an \emph{operator}, I suppose that \dots \\
\hspace{1em} \dots service $X$ at host $Y$ is up and running. \\
\hspace{1em} \dots service $X$ at host $Y$ is inaccessible longer than $Y$ min. \\
\hspace{1em} \dots network of trainee $X$ is connected to the rest of exercise \\         
    \hspace{2.4em} infrastructure. \\
\hline
\end{tabular}
\end{table}

{\bf Insight of Trainees (V\textsubscript{1})} visualizations support \textsc{trainees} in keeping track of what is happening at the moment and understanding the training content. The view on the data should be strictly person-centered and adapted to the history and performance of each particular \textsc{trainee} so that they can concentrate on the development during the training session from their perspective. 

The level of detail provided to \textsc{trainees} has to be carefully considered when designing visualizations. A visual storytelling approach to learning can provide comprehensive guidance of \textsc{trainees} throughout the training session. Using event-based visualizations emphasizing important actions and events that appeared during the \emph{execution} phase can help the \textsc{trainees} grasp the main ideas of the training content. However, this approach is rather exceptional, and visual guidance is usually intentionally restricted. A typical goal of hands-on cybersecurity training is just to exercise the \emph{perception}, \emph{comprehension}, and \emph{projection} skills of \textsc{trainees}; a subtle visual run-time support better mimics real-world conditions. The visual-based comprehension is often left for the \emph{personal feedback (V\textsubscript{4})} tools in the \emph{reflection} phase (discussed later in this Section). 

The clustering of hypotheses revealed two fields of \textsc{trainee} interest. \emph{Awareness of the state of the network environment (V\textsubscript{1A})} covers hypotheses relevant to overseeing the state of the training network maintained by a \textsc{trainee}. It is used to infer knowledge of hidden cyber events and actions from the \emph{infrastructure data (D\textsubscript{4})}. \emph{Awareness of cybersecurity posture (V\textsubscript{1B})} is related to the understanding of cyber events and actions defined as education goals in \emph{training definitions}.

{\bf Insight of Organizing Participants (V\textsubscript{2})} visualizations support \textsc{sparring partners}, \textsc{supervisors}, and \textsc{operators} in gaining insight into the state and progress of training sessions. Views are usually shared across all participants of the same role, providing them a view of the training progression, score, solved tasks, and other milestones and assessment data related to planning and timing. However, the views have to be adapted to each organizing role. V\textsubscript{2} is, therefore, divided into three categories of hypotheses according to organizing roles. \emph{Training progression (V\textsubscript{2A})} is used by \textsc{sparring partners} who need to know the current state of the \textsc{trainees'} networks and services so that they can coordinate their actions and perform them in proper order and time. \emph{Training management (V\textsubscript{2B})} of \textsc{supervisors} should be able to identify troubles of \textsc{trainees} as soon as possible. \emph{Infrastructure management (V\textsubscript{2C})} is intended for \textsc{operators} who have to monitor the unreliable infrastructure of the cyber range to detect technical problems.

Regardless of the specific role, the supervising activities of all organizing participants force them to \emph{perceive} the current state of the training, to \emph{comprehend} the situation, and to \emph{project} the future status so that the training progresses smoothly and efficiently. In contrast to the \emph{Insight of Trainees (V\textsubscript{1})}, analytical visualizations of organizing participants should fully support all these levels of awareness. 

\subsection{Visual Data Analytics}

Our classification combines user roles of the cybersecurity training life cycle (see Fig.~\ref{fig:life-cycle}) and data categories (Section~\ref{sec:data}). Table~\ref{tab:hypotheses2} summarizes the classification of hypotheses that are explained in the remainder of this section.

\begin{table}[!ht]
\centering
\caption{Visual Data Analytics: Visualization tasks V\textsubscript{3}--V\textsubscript{6} are further divided into several categories (e.g., V\textsubscript{4A}--V\textsubscript{4C}). Each category is accompanied by sample hypotheses formulated either as a prerequisite for verification (``I suppose that \dots''), or as working empirical hypothesis that is assumed to be explaining certain fact about phenomena (``I wonder \dots'' and ``I search for \dots'').}
\label{tab:hypotheses2}
\begin{tabular}{|l|}
\hline
\rowcolor{Gray}
\multicolumn{1}{|c|}{\bf V\textsubscript{3} -- Personal Feedback} \\
{\bf Personal reflection of trainees (V\textsubscript{3A}):} \\
As a \emph{trainee}, I wonder \dots \\
\hspace{1em} \dots what I did wrong in the task $X$. \\
\hspace{1em} \dots where I lost the most points and why. \\
\hspace{1em} \dots how I performed compared to other trainees.\\
{\bf Impact of supervision (V\textsubscript{3B}):} \\
As a \emph{supervisor}, I wonder \dots \\
\hspace{1em} \dots if I intervened in time. \\
\hspace{1em} \dots if I intervened properly. \\
\hspace{1em} \dots if I overlooked some troubles. \\
\hline
\rowcolor{Gray}
\multicolumn{1}{|c|}{\bf V\textsubscript{4} -- Quality of Training Exercise} \\
{\bf Correctness of a training definition (V\textsubscript{4A}):} \\
As a \emph{designer}, I suppose that \dots \\
\hspace{1em} \dots all tasks are relevant to learning objectives. \\
\hspace{1em} \dots task $X$ of the training definition $Y$ is solvable. \\
\hspace{1em} \dots the training definition $X$ is solvable as a whole (no logical \\
         \hspace{2.4em} flaws in connections and dependencies of individual tasks). \\
{\bf Difficulty of a training definition (V\textsubscript{4B}):} \\
As a \emph{designer}, I suppose that \dots \\
\hspace{1em} \dots prerequisite skills of trainees were well-defined. \\
\hspace{1em} \dots the training definition $X$ is suitable for beginners/experts/... \\
\hspace{1em} \dots teams of trainees were well-balanced \\
             \hspace{2.4em} (there were no extreme differences in skills of each team). \\
{\bf Comparison of the difficulty (V\textsubscript{4C}):} \\
As a \emph{designer}, I suppose that \dots \\
\hspace{1em} \dots the training definition $X$ is more difficult than definition $Y$. \\
\hspace{1em} \dots tasks in the training definition $X$ require more time to finish \\
\hspace{2.4em} than tasks in definition $Y$. \\
\hspace{1em} \dots assessment criteria of the training definition $X$ were \\
             \hspace{2.4em} of lower quality than assessment criteria of definition $Y$. \\
\hline
\rowcolor{Gray}
\multicolumn{1}{|c|}{\bf V\textsubscript{5} -- Behavior Analysis} \\
{\bf Successful strategies (V\textsubscript{5A}):} \\
As an \emph{analyst}, I suppose that \dots \\
\hspace{1em} \dots limiting network access is a better strategy than fixing \\
\hspace{2.4em} individual vulnerabilities in the network. \\
\hspace{1em} \dots dividing responsibility for defending individual \\
             \hspace{2.4em} hosts between team members is more efficient than \\
             \hspace{2.4em} ad-hoc defense. \\
{\bf Cooperation patterns (V\textsubscript{5B}):} \\
As an \emph{analyst}, I suppose that \dots \\
\hspace{1em} \dots closer cooperation between team members leads to more \\
             \hspace{2.4em} effective protection against attacks. \\
\hspace{1em} \dots the team $X$ had a strong leader who communicated with \\
             \hspace{2.4em} the rest of the team significantly more often. \\
\hline
\rowcolor{Gray}
\multicolumn{1}{|c|}{\bf V\textsubscript{6} -- Infrastructure Analysis} \\
{\bf Performance analysis (V\textsubscript{6A}):} \\
As an \emph{operator} or \emph{designer}, I search for \dots \\
\hspace{1em} \dots the most utilized links/nodes/CPUs \\
             \hspace{2.4em} in the infrastructure for training definition $X$. \\
\hspace{1em} \dots the peak memory usage of individual network \\
             \hspace{2.4em} nodes in training definition $X$. \\
{\bf Reliability analysis (V\textsubscript{6B}):} \\
As an \emph{operator} or \emph{designer}, I search for \dots \\
\hspace{1em} \dots the mean time to failure of nodes in the infrastructure. \\
\hspace{1em} \dots unstable custom network services in the infrastructure. \\
\hline
\end{tabular}
\end{table}

{\bf Personal Feedback (V\textsubscript{3})} to \textsc{participants} has a significant positive impact on the learning process~\cite[p.~480]{petty2009}. A good post-training visual feedback should explain the pros and cons of the chosen approach and indicate the areas for further improvement.

Effective person-centered feedback should occur as soon as possible, during or right after the \emph{execution} phase when the \textsc{trainees} remember details of their behavior, decisions, and conducted actions. Deploying such immediate visual feedback requires automated data processing and automatically generated personalized views for individual \textsc{trainees}.

Our classification scheme is divided according to roles that benefit from timely feedback: \emph{personal reflection of trainees (V\textsubscript{3A})} and \emph{impact of supervision (V\textsubscript{3B})}.

Personal feedback is crucial for the \textsc{trainees} to learn from the exercise as much as possible. Nowadays, the feedback is often restricted to providing a simple scoreboard with very limited informal comments from \textsc{supervisors} (a so-called ``hot wash-up'' session). There might be an additional debriefing later when \textsc{supervisors} manually process the data. However, the analysis is laborious, and the delayed presentation of findings might reduce the impact on \textsc{trainees}~\cite{vykopal2017}. They should receive a view of their behavior during the training session as well as comparison with other \textsc{trainees}. Moreover, the data analysis should be automated to provide in-depth feedback right after the training session. Feedback visualizations have to be well-designed and intuitive. Using common techniques would be necessary because \textsc{trainees} usually do not have time to familiarize themselves with complex tools. A low number of easy-to-decode charts (bar/line charts, scatter plots, etc.) should be favored over the complex VA tools. The user interface should motivate users to explore the data and learn from their mistakes. Applying the methods of user-centered design~\cite{norman1986,mckenna2015} is, hence, a must.

\textsc{Supervisors} can also benefit from personalized feedback after a training session since their interventions influence \textsc{trainees}. The visualizations should provide an overview as well as detailed per-trainee data. This allows \textsc{supervisors} to analyze the impact of their interventions and learn from their possible mistakes in managing the training session.

Feedback for \textsc{sparring partners} and \textsc{operators} is rare, since the main objective of the training is to teach \textsc{trainees}. This is why we omitted these two roles from the classification.

{\bf Quality of Training Exercise (V\textsubscript{4})} reflects the usefulness of training sessions for \textsc{trainees}. The main motivation is to improve future training programs by reviewing collected data by \textsc{designers}, i.e., experts with educational skills, who are responsible for the training content.
The quality can be measured and compared by various qualitative attributes that capture individual features of training sessions. \emph{Correctness}, for example, can express the ability of \textsc{trainees} to solve required tasks considering properties of the underlying infrastructure, the logical consistency of tasks, or availability of meaningful instructions. \emph{Difficulty} can be expressed as the time required to finish the training session or minimal skills required of \textsc{trainees}.
\textsc{Designers} can study either results of individual \emph{training runs} of the same \emph{training definition} or compare \emph{training definitions} mutually.

Our classification scheme divides V\textsubscript{4} hypotheses according to qualitative attributes and the multiplicity of involved training runs: \emph{Correctness of a training definition (V\textsubscript{4A})}, \emph{difficulty of a training definition (V\textsubscript{4B})}, and \emph{comparison of the difficulty (V\textsubscript{4C})}. Other qualitative attributes, apart from correctness or difficulty, can be considered. However, not all combinations are meaningful. For example, correctness typically represents a binary value (correct or incorrect) and then mutual comparison does not make sense.

The quality of a training session is primarily affected by three mutually connected factors: 
\begin{itemize}
    \item Training content defined by \emph{technical scenario (D\textsubscript{1})}. Ambiguous or illogical tasks and their extreme difficulty or simplicity can discourage \textsc{trainees} from proceeding, rendering the training session useless.
    \item Assessment defined by \emph{assessment criteria (D\textsubscript{2})}. They affect achieving educational goals. Unbalanced assessment (too lax or strict) can lead to bypassing tasks or demotivate \textsc{trainees}.
    \item Proficiency and motivation of \textsc{trainees}. The lack of knowledge, skills, or motivation can prevent \textsc{trainees} from finishing the training. Knowledge and skills are usually measured as part of prerequisite testing using questionnaires or small practical tasks.
\end{itemize}
Visual analytics can help to balance these factors by providing different views on the triplet and enabling \textsc{designers} to study their mutual interactions and dependencies so that the impact of training is maximized for a given group of \textsc{trainees}. Techniques of multiple coordinated views~\cite{roberts2007} can be used to support this exploratory analysis effectively. 

{\bf Behavior Analysis (V\textsubscript{5})} can help in discovering relevant facts about \textsc{trainees}, their skills, or behavioral patterns under stress. The observations can either reveal issues or inconsistencies in training definitions or identify general patterns applicable in practical cyber defense. For instance, visualization of users' actions can reveal patterns of successful cooperation or successful attack/defense strategies.

\emph{Successful strategies (V\textsubscript{5A})} and \emph{cooperation patterns (V\textsubscript{5B})} are two primary categories of analytical hypotheses directly related to cybersecurity education where visual perception can significantly help. The former analyzes defense and attack strategies, e.g., completely cutting off the defended network on the firewall vs. selective suspension of services being under attack. The analysis of cooperation patterns can be considered a part of the strategy analysis. However, it focuses more on people, their cooperation tactics, and how they influence the results of the training. The classification scheme can be extended to reflect other requirements of cybersecurity experts.

The raw data D\textsubscript{3} -- D\textsubscript{5} of \emph{training runs} has usually a form of time-stamped events. Reconstruction, visualization, and analysis of user processes that produced the data are possible by employ techniques of process mining~\cite{weijters2001,kriglstein2016}. Analysis of behavioral aspects can also be supported by specific statistical, knowledge discovery, or machine learning \emph{models} incorporated into the VA process (see Fig.~\ref{fig:kg-model}). For example, methods related to the node centrality in social networks~\cite{opsahl2010} can be used to identify skilled leaders in team-based training sessions. Anomaly detection algorithms~\cite{chandola2009} can identify strong/weak skills of \emph{trainees}, for instance.

These data can also serve to measure learning.~\cite{mases2017obtaining} proposes several metrics for measuring performance that are applicable in cybersecurity training. These include tracking the time spent on tasks, observing the usage of specific tools in logs, or automatically checking properties of the virtual environment, such as uptime of services. A concrete example in the context of CDXs is presented in~\cite{maennel2017improving}: the evaluators measure the time of the attack, compromise, detection, mitigation, and restoration. In~\cite{henshel2016predicting}, also non-technical aspects are measured, such as team behavior.
 
{\bf Infrastructure Analysis (V\textsubscript{6})} represents another essential activity that can affect the results and impact of cybersecurity training. Any technical difficulties or malfunctions can negatively influence \textsc{trainees}. Related visualizations should support \textsc{operators} and \textsc{designers} in exploring \emph{training definitions} and their requirements on the infrastructure and provide them with a ``backstage'' view on the operational data captured in the \emph{execution} phase.

As opposed to the \emph{infrastructure management (V\textsubscript{2C})} in situational awareness, this category relates to the feasibility of the underlying infrastructure to serve according to the prescription of the \emph{training definitions}. For example, if a heavily used server is allocated on a shared virtual node in the cyber range, then its response time can be prohibitively slow. This can hinder \textsc{trainees} in fulfilling the tasks.

Suitable visual tactics strongly depend on features and possibilities that are specific for technology used to implement the underlying infrastructure. Our classification, therefore, uses qualitative aspects that delimit generic requirements on the infrastructure: \emph{performance analysis (V\textsubscript{6A})} and \emph{reliability analysis (V\textsubscript{6B})}. The performance deals with the utilization of resources at various levels of granularity (CPU, memory, network nodes). Reliability is related to the failure rate of individual facilities. However, these two qualities represent only an example. 

\section{Demonstration} 
\label{sec:use-cases}

In this section, we illustrate the application of our conceptual model on the KYPO Cyber Range platform, which is being developed by the cybersecurity team at our university since 2013. From the beginning, KYPO was designed with an emphasis on user-friendliness and support for providing interactive visual insight into cybersecurity and learning processes. It represents a comprehensive system suitable for demonstrating the applicability of our model. As the KYPO visualizations were designed on the fly without a conceptual view towards the application domain, this section aims to demonstrate how the model fits the existing design of a complex cyber range and to reveal the undersupported parts of the training life cycle. The presented visualizations only illustrate possible approaches to the design of specific visual analysis tools.

To the best of our knowledge, other cyber ranges and cybersecurity training tools focus primarily on the training content, providing only limited visual insight. Nevertheless, we aim to discuss other approaches when the KYPO does not provide a suitable example.

\subsection{Training Life Cycles and Data in KYPO}

The \textbf{KYPO Cyber Range}~\cite{vykopal2017kypo} is a highly flexible and scalable cloud-based platform. Its core functionality is to emulate computer networks with full-fledged operating systems and network devices that mimic real-world systems. Its primary use is hands-on cybersecurity training, especially \emph{attack-only} capture the flag games and cyber defense exercises. It is also used in other cybersecurity applications, such as forensic investigation. The platform provides tools for the automated collection of various data that can be further analyzed. These include network flows, computer logs, user commands, and user actions from GUI (e.g., mouse clicks or submitted forms).

The main user interface is a web application called the \emph{KYPO portal}. We gradually extend the set of available visualizations and visual analytics tools integrated into the \emph{KYPO portal} using the participatory design process. Nine cybersecurity experts (two specializing in cybersecurity education who are co-authors of this paper) closely collaborated in the design and evaluation of novel visualizations and the improvement of their features.

\textbf{Capture the Flag} games consist of tasks divided into consecutive levels where access to the next level is conditioned by completing the previous one. Players can use hints or skip entire levels. These actions (taking hints and skipping or completing a level) are penalized or rewarded by scoring points. The final scores of individual \textsc{trainees} within the same session are mutually comparable and can be used for their evaluation. A typical session lasts for one to two hours. Several \textsc{supervisors} facilitate a group of up to 20 \textsc{trainees} working as individuals or in pairs.

\textsc{Designers} of CTF games are experts from the cybersecurity incident response team of our university or undergraduate students of a one-semester course on designing cybersecurity games~\cite{svabensky2018}. They produce \emph{training definitions} that describe both \emph{technical scenarios (D\textsubscript{1})} and \emph{assessment criteria (D\textsubscript{2})}. The training definition is a set of (plain text) documents that include: a description of the network environment and the configuration of individual network nodes (including vulnerabilities to be exploited in the game levels); a common background story and task descriptions (for each level); definition of hints, worked-out solutions and penalty points for taking hints (for each level); the \textsc{trainee}'s prerequisites, educational objectives and further assessment criteria. \emph{Designers} can interactively prepare content and allocate resources required for training sessions through the KYPO portal.

The produced \emph{training definitions} are used for creating training sessions in the \emph{execution} phase. The KYPO Cyber Range automatically logs \textsc{trainees}' \emph{user actions (D\textsubscript{3})}. Some of the \emph{training definitions} contain pre- and post-game questionnaires for assessing \textsc{trainee} knowledge (i.e., \emph{assessment data (D\textsubscript{5})}), which is stored as well. So far, \emph{infrastructure data (D\textsubscript{4})} collection is not supported in CTF games.

\textbf{Cyber Czech} is a series of technical cyber defense exercises for up to six \emph{blue teams} (3--4 members). The \textsc{trainees} must protect their infrastructure against various attacks from the \emph{red team} and fulfill requests from other \textsc{sparring partners}, as defined in Sec.~\ref{sec:roles}. The exercise spans two days. During the first day, the \textsc{trainees} familiarize themselves with the virtual environment. The second day is devoted to the actual training session, which lasts 6 hours. A brief (up to 30 minutes) personalized feedback session follows right after the exercise. Finally, there is another feedback session approximately two weeks later, in which organizers elaborate on the strengths and weaknesses of each team. From each exercise, we collect network flows, computer logs, user commands, and automatic and manual scoring records.

The variability and complexity of CDXs are substantially bigger than in CTFs. The preparation of a new training run of Cyber Czech exercise takes tens of person-months. A unique training definition is created almost from scratch each year and is only repeated a few times. Only a GUI for the \emph{execution} and \emph{reflection} phases are currently supported in the KYPO Portal, both to a limited extent.

The \emph{technical scenario (D\textsubscript{1})} is comprised of the infrastructure of nearly 200 computer nodes in multiple local networks, scheduled attacks and respective vulnerabilities, and configuration of monitoring tools for both trainees and organizers. Multiple iterations make the preparation very laborious. Each Cyber Czech exercise series is framed with a unique story and additional non-technical tasks. The \emph{assessment criteria (D\textsubscript{2})} include several dozen automatically scored network services (e.g., availability of web server or database) and up to 30 manually scored tasks (e.g., penalties for individual attacks, communication with the \textsc{sparring partners} from the \emph{white} team or \emph{fictitious users}), and requests for reverting malfunctioned network nodes. Complex dependencies in which one network service (e.g., active directory) depends on other services (such as DNS) often exist. All this complicates the design and implementation of a unified data scheme and corresponding front-end tools. Correctness and the estimation of difficulty of training definitions are addressed by so-called ``dry runs'' in which the whole exercise is tested by volunteers. However, the approach is costly and can be misleading because the readiness of testers may significantly differ from the readiness of target learners.

\subsection{Visual Analytics of Capture the Flag Games}

\begin{figure}[!ht]
  \centering
  \includegraphics[width=\columnwidth]{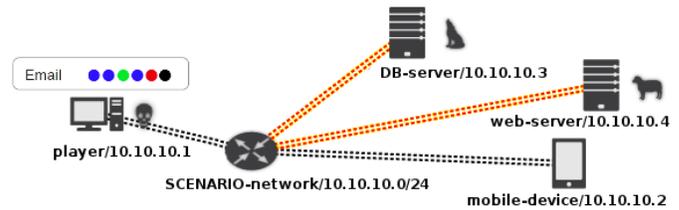}
  \caption{Network Topology with glyphs supporting situational awareness.\protect\footnotemark[1]}
  \label{fig:topology}
\end{figure}

\textbf{Insight of Trainees (V\textsubscript{1}).}
\textsc{Trainees} gain insight into the game content through the web-based KYPO portal, which provides them with task descriptions, hints, and solutions for each level and also shows information about the current level and remaining time of the training session. The \emph{Network Topology} visualization (Fig.~\ref{fig:topology}) mediates remote access to individual hosts via a web browser and provides situational awareness by decorating a simple network graph with various semantic symbols. For example, it is possible to support V\textsubscript{1A} by coloring network links depending on current throughput, and V\textsubscript{1B} by glyphs distinguishing logical roles of hosts (attacker, victim), or events captured in hosts (e.g., received mails). The importance and quantity of this semantic data differ between training definitions, and they also vary in time. Combining them meaningfully and showing them at the right time so that the \textsc{trainees} are not overburdened is a challenging task.

\footnotetext[1]{We provide a full-page version of the visualization in Supplementary Materials at \url{https://www.kypo.cz/media/3197111/tvcg19-supplemental-materials.pdf}}

\textbf{Insight of Organizing Participants (V\textsubscript{2}).}
Since we currently support attack-only CTFs without \textsc{sparring partners}, no special visualizations for V\textsubscript{2A} exist in KYPO.

\begin{figure}[!ht]
  \centering
  \includegraphics[width=\columnwidth]{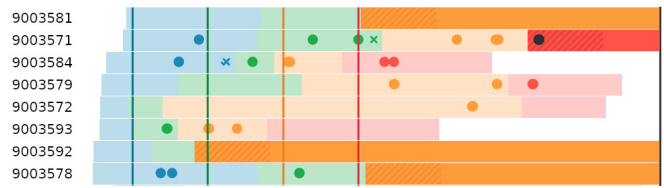}
  \caption{CTF Training Session Overview shows the progress of individual trainees during the training session.\protect\footnotemark[1]}
  \label{fig:hurdling}
\end{figure}

\textsc{Supervisors} use \emph{CTF Training Session Overview} visualization (Fig.~\ref{fig:hurdling}) that displays the progress of \textsc{trainees} throughout the CTF game. Each row captures the training session of individual \textsc{trainees}, who can start at slightly different times. Colored bars represent levels. Dots represent user events (e.g., taking a hint), vertical lines show expected level duration. \textsc{Supervisors} use this view to actively manage the training session (V\textsubscript{2B}) by looking for \textsc{trainees} in trouble (e.g., those stuck in a level for too long, those repeatedly trying to guess the flag to pass the level instead of solving the task, or those about to quit without trying, which is signaled by displaying all the hints and the solution shortly after each other). 

Since our CTFs are executed in the complex cloud-based KYPO Cyber Range, dealing with technical issues is delegated to specialized \emph{operators} managing this infrastructure. They gain insight into the infrastructure state (V\textsubscript{2C}) via off-the-shelf OpenNebula Sunstone dashboard (see supplemental materials\protect\footnotemark[1]).

\begin{figure*}[!ht]
  \centering
  \includegraphics[width=\textwidth]{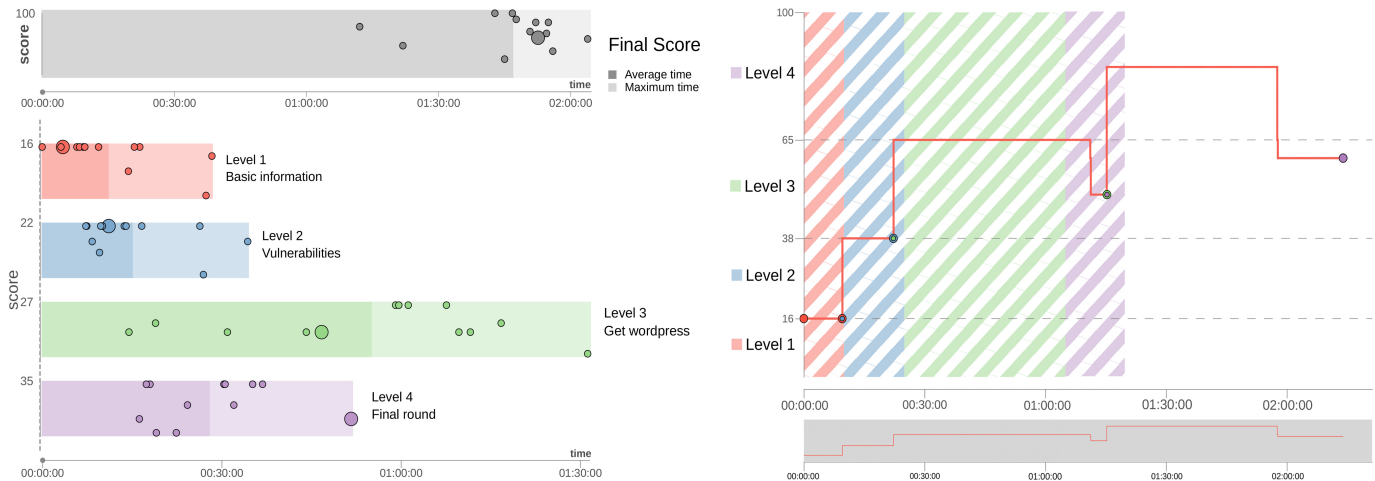}
  \caption{CTF Feedback Dashboard providing individual view on \textsc{trainee}'s score results and development in time.\protect\footnotemark[1]}
  \label{fig:ctf-feedback-dashboard}
\end{figure*}

\textbf{Personal Feedback (V\textsubscript{3}).}
At the end of a session, \textsc{trainees} receive a \emph{CTF Feedback Dashboard}~\cite{oslejsek2019} supporting V\textsubscript{3A} with two complementary views (Fig.~\ref{fig:ctf-feedback-dashboard}). The left view provides the final score overview for comparison with other \textsc{trainees}. The lengths of the bars show the time of the slowest trainee; different color intensity provides information about the average time. The right side of the dashboard displays the individual score development in time throughout the game. The width of striped areas represents time spent in levels. Dots represent user events. A very similar dashboard is used by \textsc{supervisors} (\emph{V\textsubscript{3B}}) who, in addition, can plot multiple \textsc{trainees} into the score development time series chart for comparison.

\textbf{Quality of Training Exercise (V\textsubscript{4}).}
Qualitative aspects of CTF \emph{training definitions} are supported in KYPO by simple statistical visualizations, e.g., histograms and boxplots capturing the distribution of scores gained by \textsc{trainees}. The \emph{CTF Feedback Dashboard} (Fig.~\ref{fig:ctf-feedback-dashboard}) from \emph{personal feedback (V\textsubscript{3})} can be also used to identify weak parts of the training, e.g. levels where \textsc{trainees} spend a long time. However, deeper research and the design of narrowly focused visualizations for quality-related analysis is a future work opportunity.

\textbf{Behavior Analysis (V\textsubscript{5}).}
Behavior in connection with cybersecurity is often linked to attack graphs and estimation of weak points in the network. A study~\cite{bassett2016} introduced a method for analyzing computer network security. The method operates with attack paths that represent a linkage of individual nodes with conditions of compromised network security. The output is an attack graph with behavior prediction, and the authors propose the use of their method for incident response training. As for CTF games, the method could also bring insight to the trainee's actions and help the instructor to monitor progress or strategies.

\textbf{Infrastructure Analysis (V\textsubscript{6}).}
The already mentioned off-the-shelf dashboard provided by OpenNebula Sunstone is currently used also for the basic qualitative evaluation of the underlying cloud infrastructure of the KYPO Cyber Range. However, its utilization for these tasks is not very effective, as it is a universal cloud management tool.

\subsection{Visual Analytics of Cyber Czech}

\begin{figure}[!ht]
  \centering
  \includegraphics[width=\columnwidth]{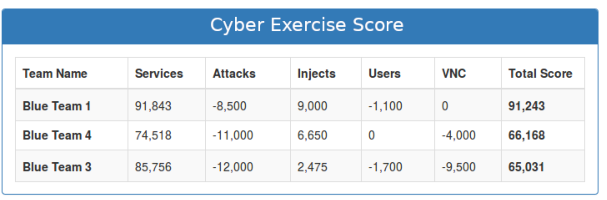}
  \caption{CDX Scoreboard shows the current scores of all \emph{blue teams}.\protect\footnotemark[1]}
  \label{fig:cdx-scoreboard}
\end{figure}

\textbf{Insight of Trainees (V\textsubscript{1}).}
Since Cyber Czech is mainly a technical exercise, awareness of the network state \emph{V\textsubscript{1A}} and cybersecurity posture \emph{V\textsubscript{1B}} are intentionally restricted to resemble real-world settings, as discussed in Section~\ref{sec:viz}. \textsc{Trainees} interact with a network topology visualization similar to Fig.~\ref{fig:topology}. However, the network infrastructure is more complex, and there are no semantic decorations. Instead, the \textsc{trainees} use a standard monitoring tool (Nagios) showing the status of the network services they are trying to protect. Further, they can infer the consequences of their actions only from the real-time \emph{CDX Scoreboard} (Fig.~\ref{fig:cdx-scoreboard}) displayed during the exercise. The scoreboard shows the current total score as well as per-category scores and penalties of all \emph{blue teams}, allowing them to compare themselves. The use of a restricted table-based view is intentional, as we aim to simulate real conditions during the CDX with only limited real-time feedback.

\begin{figure}[!ht]
  \centering
  \includegraphics[width=\columnwidth]{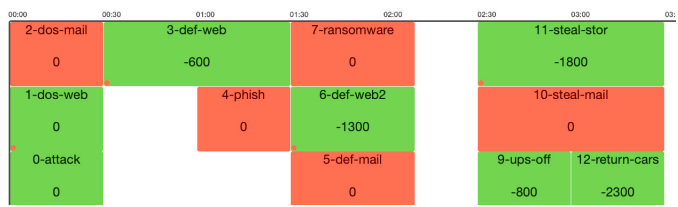}
  \caption{CDX Attack Plan displays scheduled attacks of the \emph{red team} at the end of a 6-hour long training session.\protect\footnotemark[1]}
  \label{fig:cdx-attack-plan}
\end{figure}

\textbf{Insight of Organizing Participants (V\textsubscript{2}).}
\emph{Training progression (V\textsubscript{2A})} of the \emph{red team} is supported by \emph{CDX Attack Plan} (Fig.~\ref{fig:cdx-attack-plan}) showing the interactive plan of individual attacks and their state (inactive/ongoing/completed). The green color stands for successful attacks; red stands for unsuccessful ones (i.e., the \emph{blue team} has defended themselves). Attack type abbreviations and given penalty points are shown within each block. Clicking on an attack block reveals further details (e.g., additional comments or screenshots).
The \emph{green team} uses the \emph{Nagios} service monitoring system to watch the infrastructure (\emph{V\textsubscript{2C}}), to detect when the trainees (un)intentionally blocked some of the monitored and scored services, and to provide brief advice (\emph{V\textsubscript{2B}}). 
Visual insight of other organizing participants is not currently supported.

\begin{figure}[!ht]
  \centering
  \includegraphics[width=\columnwidth]{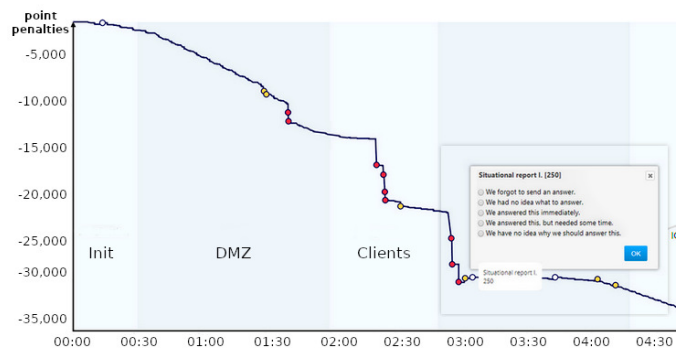}
  \caption{CDX Personalized Feedback shows the score development throughout the training session of a single blue team.\protect\footnotemark[1]}
  \label{fig:cdx-feedback}
\end{figure}

\textbf{Personal Feedback (V\textsubscript{3}).}
During the hot-washup session, organizers give immediate verbal feedback to \textsc{trainees}. \emph{Personal reflections on the trainees (V\textsubscript{3A})} are supported by presenting them the \emph{CDX Attack Plan} (Fig.~\ref{fig:cdx-attack-plan}) that was hidden from the \textsc{trainees} during the exercise. \textsc{Trainees} are also provided with the \emph{CDX Personalized Feedback}~\cite{vykopal2018} (Fig.~\ref{fig:cdx-feedback}) that shows the score development of their \emph{blue team}. Dots include details about penalties entered by \emph{red}, \emph{white}, and \emph{green teams}. Each dot is associated with a short feedback poll used for gathering further information from \textsc{trainees}. The data is used in the follow-up analysis.
The \emph{impact of supervision V\textsubscript{3B}} is not currently supported.

\textbf{Quality of Training Exercise (V\textsubscript{4}).}
Vorobkalov and Kamaev \cite{vorobkalov2008} describe an approach to the quality estimation of e-learning systems. Their learning process model is based on an extended stochastic Petri net. The method has been implemented in an automated system, and it focuses on helping the expert to perform e-learning process analysis and to deduce learning course mistakes. However, it covers only systems based on net models. For CDX training, the model would not reflect the closely related state of the operational environment. Furthermore, when we consider the unstructured nature of CDX, the model would have to be very sophisticated and extensive. 

\textbf{Behavior Analysis (V\textsubscript{5}).}
The above-mentioned method by Bassett and Gabriel \cite{bassett2016} can also be applied to the CDX use case. In this embodiment, the method could be utilized in the form of an attack tool to execute or simulate the events and conditions in the attack graph. The trainee would then receive the output, helping them identify attacks they were facing and allowing them to learn from the events retrospectively (since in CDX, we don't usually want to give them any instant feedback). However, such output would have to be further transformed into a visual form suitable for this type of training.

\begin{figure}[!ht]
  \centering
  \includegraphics[width=\columnwidth]{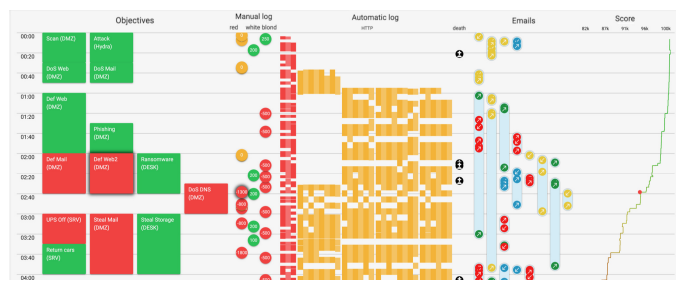}
  \caption{Prototype of CDX Analytical Dashboard.\protect\footnotemark[1]}
  \label{fig:cdx-analytical-dashboard}
\end{figure}

\textbf{Infrastructure Analysis (V\textsubscript{6}).}
The support for this type of visual analysis is essentially non-existent at the moment. Although the KYPO platform collects some types of relevant data (e.g., system logs and commands entered by blue teams at individual network nodes), the data is processed ad-hoc and manually or not at all. This is usually done for a debriefing meeting of the organizing participants about a week after the training session. The attendees summarize their observations backed by collected data (e.g., feedback forms from the \textsc{trainees}, analysis of the score development). To support the discussion, we are developing an analytical tool for CDX evaluation that will provide a timeline visualization of automatic and manual logs together with the communication threads among the \emph{blue team} and corresponding \emph{white team} members (Fig.~\ref{fig:cdx-analytical-dashboard}).

\section{Discussion}
\label{sec:discussion}

In this section, we emphasize four key observations we attained and present the challenges for future visualization research in the domain.

\textit{The current visualization tools support only situational awareness during the execution phase.} The main focus of training sessions is on the execution phase. Therefore, visualizations are designed to provide insight both to trainees (V\textsubscript{1}) and organizing participants (V\textsubscript{2}). The reflection phase, in contrast, is vastly unsupported, with the exception of personal feedback (V\textsubscript{3}) for trainees.  

\textit{Organizers have limited insight into the educational impact on learners.} The design of cybersecurity training sessions is driven mainly by technical aspects. Training sessions often aim at mastering a particular cybersecurity technique or procedure without focusing on broader learning goals. To overcome this issue, the top-down approach of designing the training must be applied, starting from defining learning goals and going down to a selection of particular techniques. Visual measuring and comparing the quality of learned skills, which is largely overlooked, could help in this process. There is a broad unexplored research area in training quality (V\textsubscript{4}) and behavior (V\textsubscript{5}) analysis.

\textit{Organizers underestimate infrastructure monitoring and analysis.} CTF and CDX depend heavily on customized monitoring and management tools for the underlying infrastructure (V\textsubscript{2C}). However, these tools are lacking. Low-level monitoring tools and other general-purpose solutions, which do not provide a complex overview of the situation, are preferred to customized ones. Analytical tools for post-event infrastructure analysis (V\textsubscript{6}) are also lacking.

\textit{Data collection is not a problem; data processing is.} It is possible to collect large amounts of multivariate data either from the emulated network environment (e.g., network flows, computer logs, commands entered) or from the user interfaces of the cyber range (e.g., mouse tracking, and clicks). The bottleneck lies in data processing and presentation, as we point out in the demonstrative examples. Especially in CDX, data correlation is a difficult task. With rising interest in the quality of training exercise (V\textsubscript{4}), a behavior analysis (V\textsubscript{5}) could accelerate the demands on the use of the data.

\begin{table}[!ht] 
\centering
\caption{The mapping of the low-level roles on the visualization tasks.}
\label{tab:roles-va}
\begin{tabular}{rccccc}
    & trainee & \makecell{sparring\\ partner} & supervisor & designer & operator \\
    \rowcolor{Gray} V\textsubscript{1} & \textbullet & & & & \\
                    V\textsubscript{2} & & \textbullet & \textbullet & & \textbullet \\
    \rowcolor{Gray} V\textsubscript{3} & \textbullet & & \textbullet & & \\
                    V\textsubscript{4} & & & & \textbullet & \\
    \rowcolor{Gray} V\textsubscript{5} & \textbullet & \textbullet & \textbullet & \textbullet & \textbullet \\
                    V\textsubscript{6} & & & & \textbullet & \textbullet \\
\end{tabular}
\end{table}

Challenges for the visualization community are a reflection of the absence of tools. Table~\ref{tab:roles-va} summarizes users who benefit from the six visualization tasks, as revealed by the conceptual model in Section~\ref{sec:viz}. Each bullet represents a visually-analytical use case.
However, only a few use cases are somehow covered in current practice. For the post-exercise analysis, the main challenge is to find meaningful uses of the collected data to improve the \textsc{supervisors}' understanding of \textsc{trainees} skill development as well as to provide insight into the training processes for \textsc{designers}. Another challenge is to design and develop VA tools to help the \textsc{designers} and \textsc{organizers} test their hypotheses. Last but not least, it is necessary to revisit the tools for situational awareness of participants during the exercise and provide them with timely individual feedback.

\section{Conclusion and Future Work}
\label{sec:conclusion}

Hands-on cybersecurity training is crucial in educating the future workforce. However, measuring the effectiveness of the training process, using either technical or educational indicators, remains largely unexplored. Our work is motivated by a desire to improve these aspects by applying visual analytics. To the best of our knowledge, this paper is the first attempt to describe the application of VA models to hands-on cybersecurity education.

We used software engineering methods to describe the training life cycle and formalize user roles involved in cybersecurity training sessions. The foundations of our work lie in the existing generic VA models. We systematized the visualizations and hypotheses into six categories and demonstrated the application of the VA model on two classes of cybersecurity training hosted at the KYPO Cyber Range platform.
The main limitation is the lack of details from other cyber ranges and training sessions. However, we assume that they are on a similar level of maturity. We back this claim with the experience of our university cybersecurity team members from their participation in events similar to the Cyber Czech exercise series.

Each of the six visualization tasks of the presented conceptual model deserves further investigation. The definition of specific guidelines that can help VA designers and researchers build visual tools is out of the scope of this paper. However, this paper aims to serve as a framework for such guidelines, providing researchers relevant use cases where the application of VA is demanding. We hope that our work will help to establish the agenda for advancing the state of the art and motivate other visualization researchers to explore the domain in which tehe research areas of education, cybersecurity, and data visualization intersect.

\section*{Acknowledgment}

This research was supported by ERDF ``CyberSecurity, CyberCrime and Critical Information Infrastructures Center of Excellence'' (No. CZ.02.1.01/0.0/0.0/16\_019/0000822).
Computational resources were provided by the European Regional Development Fund Project CERIT Scientific Cloud (No. CZ.02.1.01/0.0/0.0/16\_013/0001802).

\IEEEtriggeratref{39}
\bibliographystyle{IEEEtran}
\bibliography{IEEEabrv,tvcg2020.bib}

\newpage

\begin{IEEEbiography}[{\includegraphics[width=1in,height=1.25in,clip,keepaspectratio]{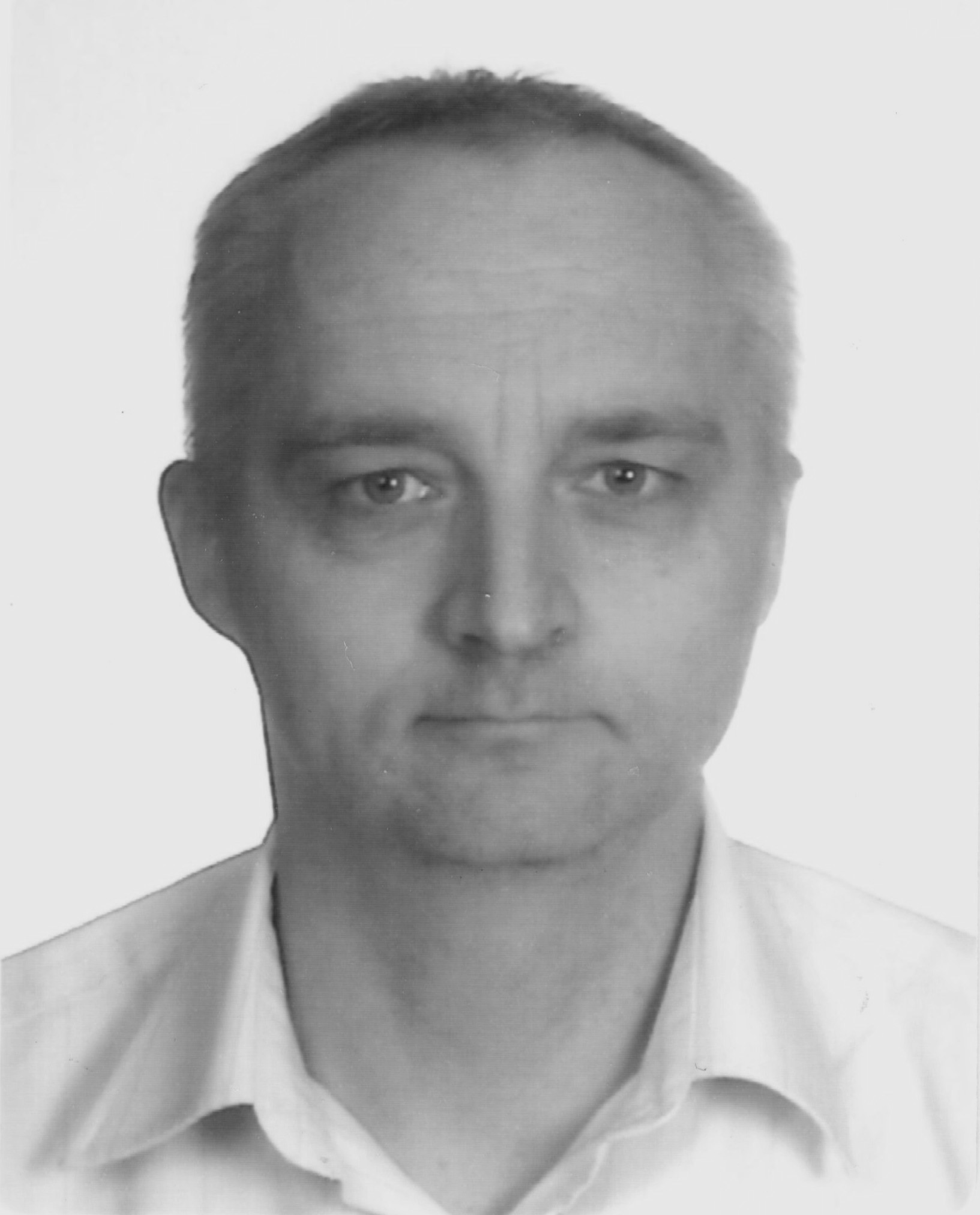}}]{Radek Ošlejšek} 
received his Ph.D. degree in informatics from Masaryk University in Brno, the Czech Republic, in 2004 for the application of software engineering methods to the area of computer graphics. He is an assistant professor with the Faculty of Informatics, MU Brno. His current research interests include interactive visualizations, knowledge modeling, and exploratory data analysis.
\end{IEEEbiography}

\begin{IEEEbiography}[{\includegraphics[width=1in,height=1.25in,clip,keepaspectratio]{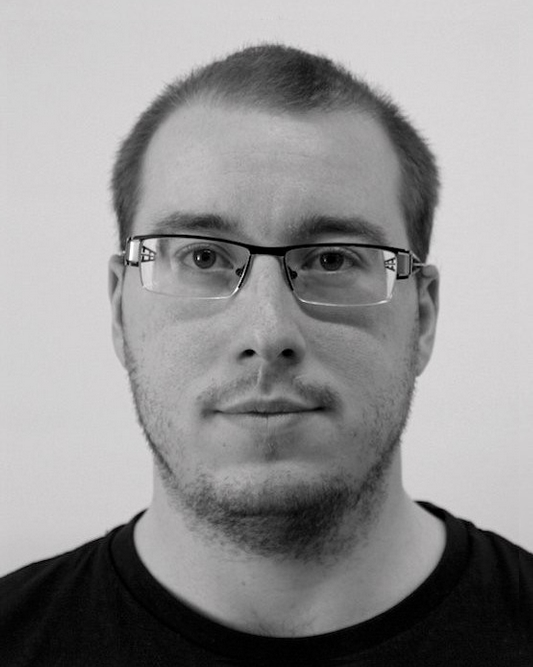}}]{Vít Rusňák}
is a researcher at the Computer Security Incident Response Team at Institute of Computer Science, Masaryk University. He received a PhD degree in Informatics from Masaryk University in Brno, the Czech Republic in 2016. His research interests include the user-centered design of interactive visualizations and collaborative user interfaces.
\end{IEEEbiography}

\begin{IEEEbiography}[{\includegraphics[width=1in,height=1.25in,clip,keepaspectratio]{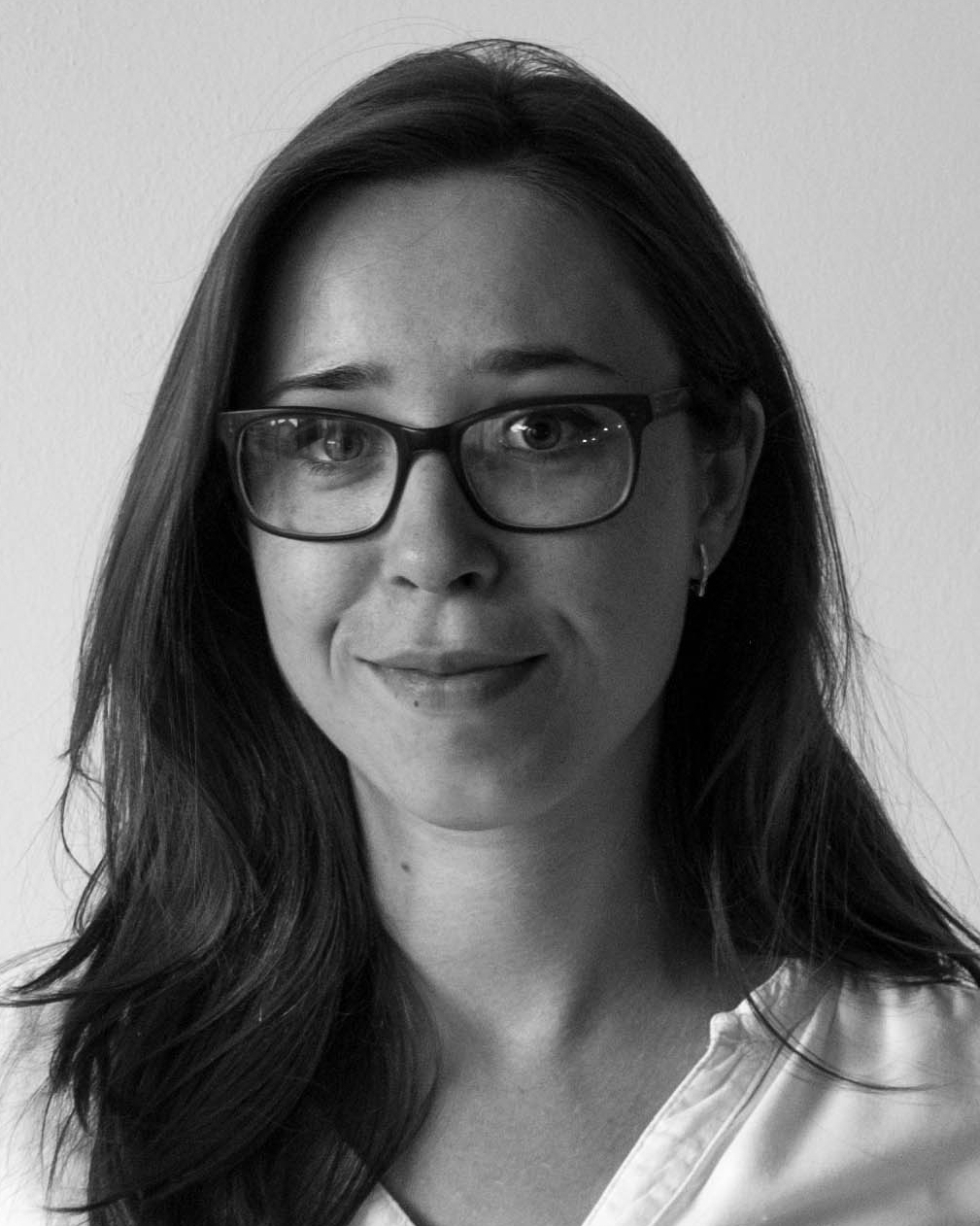}}]{Karolína Burská}
is currently a PhD student of computer science at Masaryk University in the Czech Republic. In her research, she aims at visualization in the context of cybersecurity education. As a member of a team of Masaryk University called KYPO, which focuses on simulation and mitigation of cybernetic threats, she focuses on interactive techniques in scientific visualization and exploratory analytics within cybersecurity.
\end{IEEEbiography}

\vfill

\newpage

\begin{IEEEbiography}[{\includegraphics[width=1in,height=1.25in,clip,keepaspectratio]{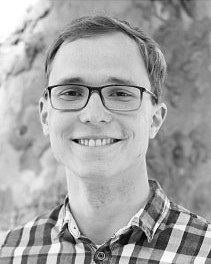}}]{Valdemar Švábenský}
enjoys teaching, so it is no surprise that he researches how to train new cybersecurity experts effectively. Specifically, he analyzes data from KYPO cybersecurity games to provide personalized feedback to learners who practice their offensive security skills. He actively participates in computing education conferences and received the Masaryk University award for the best teachers.
\end{IEEEbiography}

\begin{IEEEbiography}[{\includegraphics[width=1in,height=1.25in,clip,keepaspectratio]{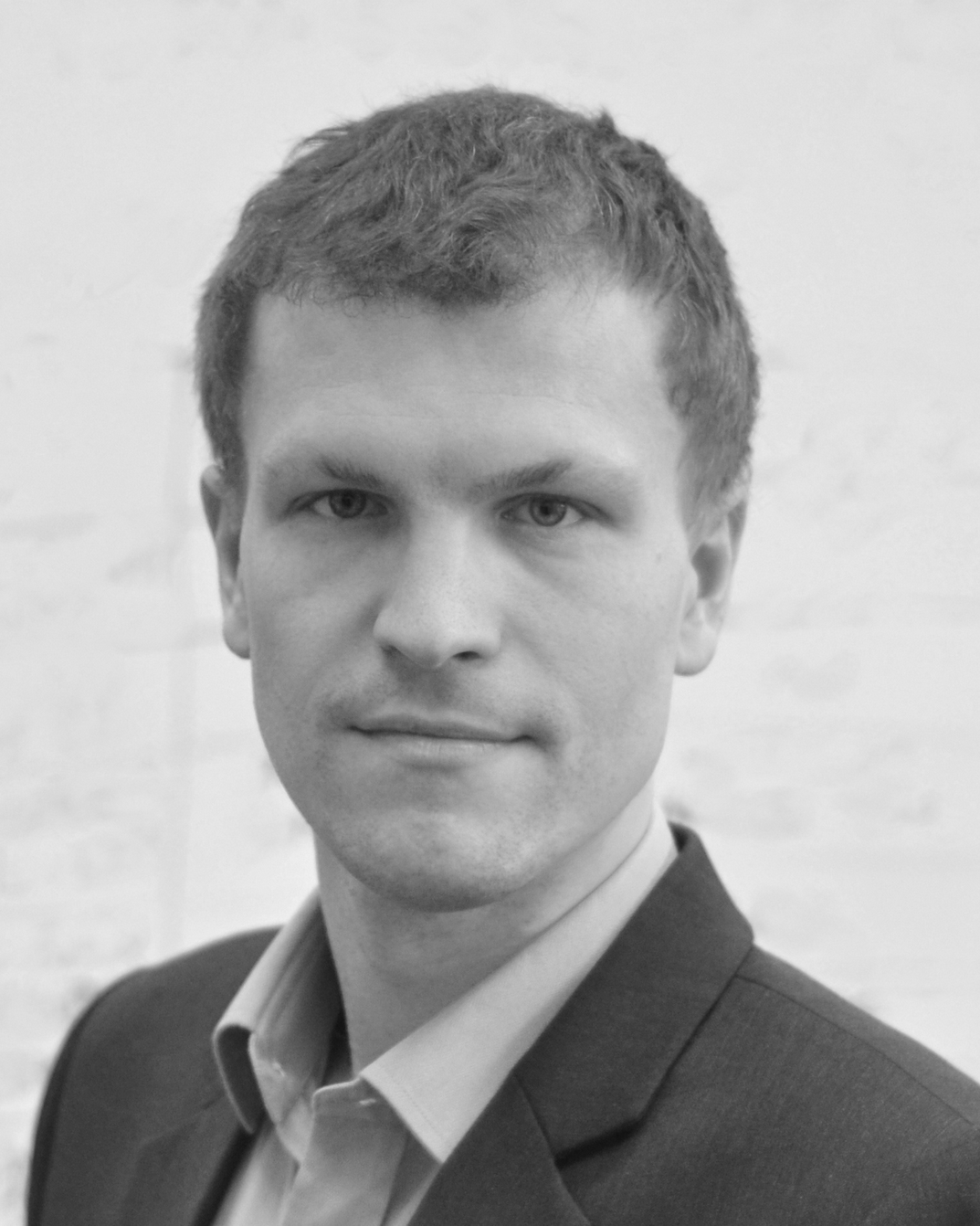}}]{Jan Vykopal}
received the PhD degree from Masaryk University, Brno, in computer systems and technologies in 2013 for network-based intrusion detection in high-speed networks. His current research interest is cybersecurity education, particularly active learning using cyber ranges and virtual environments. Jan has been designing and organizing various cybersecurity games and exercises, including the Czech national defense exercise, since 2015.
\end{IEEEbiography}

\begin{IEEEbiography}[{\includegraphics[width=1in,height=1.25in,clip,keepaspectratio]{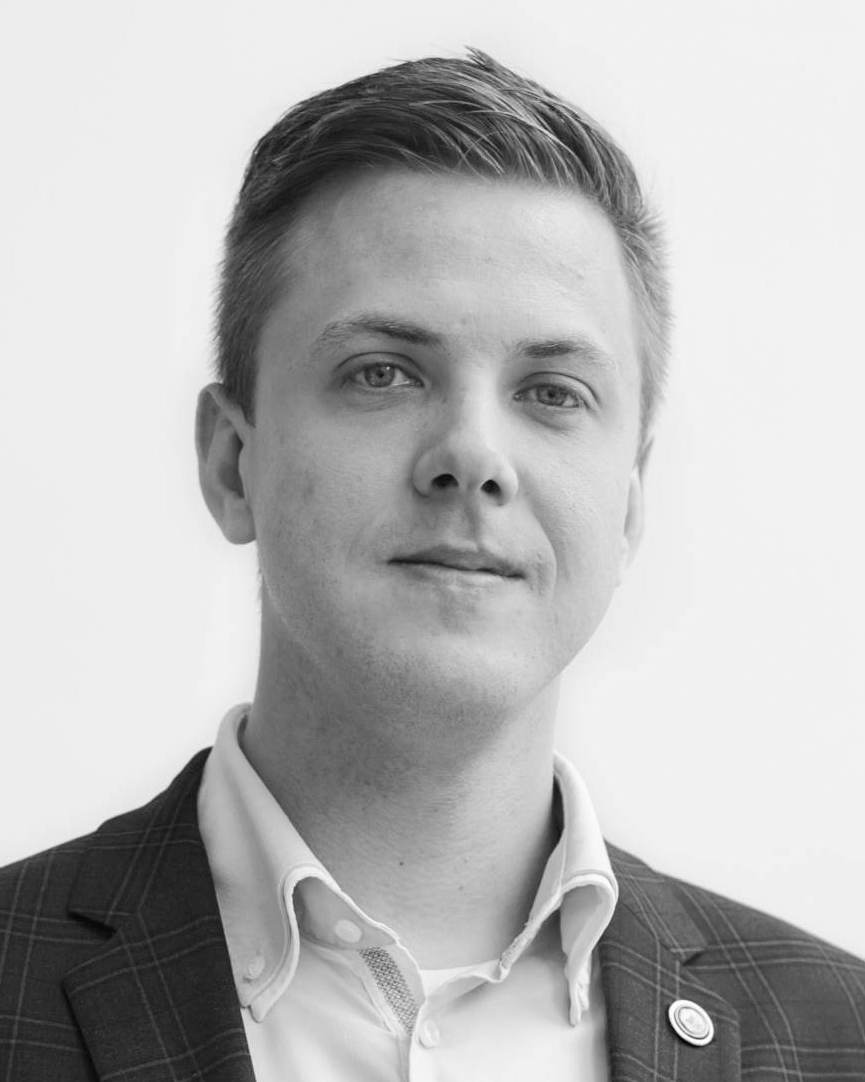}}]{Jakub Čegan}
is KYPO Cyber Range Platform and Cyber Defence Exercise (CDX) project manager. His area of interest is the development of meaningful and engaging CDX and training and providing them to customers.
\end{IEEEbiography}

\vfill

\end{document}